# Observation of the smallest metal nanotube with square-cross-section

M.J. Lagos[1,2], F. Sato[2], J. Bettini[1], V. Rodrigues[2], D.S. Galvão[2], D. Ugarte [1,2]*

[1]Laboratório Nacional de Luz Síncrotron, C.P. 6192,
13083-970 Campinas SP, Brazil.

[2]Instituto de Física "Gleb Wataghin", Universidade Estadual de Campinas-UNICAMP,
13083-970 Campinas SP, Brazil.

*To whom correspondence should be addressed. E-mail: dmugarte@ifi.unicamp.br

**Understanding the mechanical properties of nanoscale systems requires a range of measurement techniques and theoretical approaches to gather the relevant physical and chemical information. The arrangements of atoms in nanostructures and macroscopic matter can be different, principally due to the role of surface energy, but the interplay between atomic and electronic structure in association with applied mechanical stress can also lead to surprising differences. For example, metastable structures such as suspended chains of atoms[1-3] and helical wires[4,5] have been produced by the stretching of metal junctions. Here we report the spontaneous formation of the smallest possible metal nanotube with a square cross-section during the elongation of silver nanocontacts. Ab initio calculations and molecular simulations indicate that the hollow wire forms because this configuration allows the surface energy to be minimized, and also generates a soft structure capable of absorbing a huge tensile deformation.**

The study of metal junctions has mainly been associated with gold nanowires[6] due to the easiness of performing experiments with a low reactivity metal. Considering other noble metals, silver shows a face centered cubic (fcc) structure with almost identical lattice parameter of Au, but subtle changes of the surface energy (cubic facets {100} gain importance over the compact {111} planes[7]) generate clear differences in structural and mechanical behavior in comparison with Au nanowires[8,9]. Fig. 1 shows the stretching of a silver nanowire along [001] direction (hereafter noted [001] nanowire) (Supplementary Material, Video01). Initially, the wire shows a rod-like morphology and, gradually thin to finally form a linear atom chain. The evolution of Ag suspended atom chains is rather fast (see discussion in Methods Section) what renders difficult a clear identification of the atomic positions and precise measurement of bond lengths. The estimatives from intensity profile curves suggest bonds values of 3.4-3.6 Å (image 10.3 s in Fig. 1).

As {100} facets are favored in silver nanoparticles[7,8], it is expected that [001] nanorods would display square cross-section. At 0 s, the atomically resolved image shows a flattened hexagonal pattern, that agrees with a square fcc Ag rod whose width is one lattice parameter ($a_0$ = 4.09 Å) and, being observed along the [110] direction (Fig. 2a, left side). The 3.6 s image shows a bamboo-like contrast pattern formed by squares of side $a_0$ and, a rather bright center. The observation of this kind of contrast in [001] Ag nanowires is not a rare event. We have observed many different silver rods with bamboo-like-contrast obtained from different experimental conditions (room and low temperature experiments, see Supplementary Material, Fig. S1).

As a HRTEM image showing square symmetry can only occur for observations along cubic axes (or <100> kind of directions, Fig. 2a left side), a rotation of 45 degrees ([110] $\Rightarrow$ [100]) around the nanowire [001] axis must be considered to explain the image change from a flattened hexagonal- to a square bamboo-like contrast pattern. Nevertheless, we must note that we have not observed any axial rotation of nanowires for other studied systems (ex. Au[9], Ag[8,10], Cu[11]). Surprisingly, we have noted that when a nanowire displays a bamboo-like image, a fluctuation between bamboo-like contrast and deformed hexagonal pattern occurs during nanowire elongation (Fig. 3a).

We must emphasize that a one-lattice-parameter-wide [001] fcc wire observed along [100] axis does not account for the bamboo-like contrast, because a square arrangement of periodicity $a_0/2$ should be observed (Fig. 2a, left side). To explain the bamboo-like contrast pattern, we must simply remove the atoms at the wire axis (grey colored atoms in Fig. 2a), what generates a hollow square cross-section nanowire. To test this model, we can perform a more detailed analysis of HRTEM images. For example, a close look at the 3.6 s micrograph (Fig. 1) shows that the black disks (indicating atom positions) located at the center of the square sides are darker that the ones at the corners. HRTEM images represent a bi-dimensional projection of the atomic structure and, contrast of atomic columns can be, in a first approximation, considered proportional to the number of atoms along the observation direction[10,12]; this provides a way to test both projected and three dimensional proposed atomic arrangement. The comparison between intensity profiles extracted from the bamboo-like contrast micrograph and simulated HRTEM image of a hollow square Ag pillar yields excellent agreement (Fig. 2b), strongly supporting the interpretation by means of a tubular wire structure.

One important issue to be addressed is whether the tubular structure stability could be the result of the presence of light elements or contaminants invisible to HRTEM. Extensive *ab initio* calculations allow us to rule out this possibility. Details of this investigation will be published elsewhere.

The suggested hollow [001] nanowires are formed by the stacking of two different planes containing 4 atoms each ($4_A/4_B$ stacking), instead of the 5/4 stacking in a fcc wire (Fig. 2a). A tubular wire formed by the stacking of two atomic planes containing both 4 atoms, should display two energetically equivalent atomic arrangements obtained by just interchanging the A,B sites along the wire. It is very likely that these two energy minima should present a rather low energy barrier mode, in such a way that the wire should fluctuate between these two configurations. In structural terms, this process may occur by contracting the planes $4_A$ to generate $4_B$ one, while producing the opposite at the $4_B$ plane (expansion $\rightarrow 4_A$). When looked in a HRTEM experiment, the suggested process (breathing-like) would appear as a nanowire axial rotation exchanging [100] and [110] observation directions (see Fig. 3 and, Supplementary Material Video02). As a matter of fact, this structural fluctuation is indeed observed (apparent nanowire axial rotation discussed above, Fig. 3) and, then, it provides unambiguous confirmation of the hollow nanowire structure formation.

We have analyzed the structural stability by total energy calculations of Ag nanowires using *ab initio* Density Functional Theory (DFT) methodology[13,14]. Fcc wires formed by 5 up to 13 atomic planes (2-6 $a_0$ in length) were first analyzed. The layers at the nanowire ends always contained 5 atoms and are kept frozen during the geometrical optimizations in order to simulate the lattice matching of a perfect crystal; the corresponding tubular wires were formed by introducing 1 up to 5 vacancies in the fcc wire axis. The related formation energy by atom of the relaxed structures is shown in Table 1; Fig. 4a shows an example of the derived relaxed atomic wires. The calculations indicate that tubular nanowires are slightly less stable than fcc ones, this implies that the elastic energy associated with the elongation is contributing for the formation and stabilization of these peculiar hollow nanowires. Relaxed tubular structures show a slight corrugation of the lateral [100] type of facets due to the contraction of atomic planes where vacancies were introduced. This corrugation has not been observed experimentally, probably due to the typical experimental error ~0.2 Å for distance measurements from the time-resolved HRTEM observations.

A question which naturally arises is how the stretching induces a structural evolution leading to the metastable Ag nanotube formation. In this sense, it must be mentioned that several theoretical groups have reported extended search for the Ag nanowire structure generated during the wire thinning processes[15,16], but hollow structures have not been considered. Then, we have investigated the Ag wire elongation using molecular simulations; firstly, we have analyzed wether a one-lattice-parameter-wide fcc Ag nanowire (5/4 stacking) can evolve to form the hollow wires. We have taken the relaxed fcc wire formed by 13 layers from the previous calculations (total length 24.082 Å) and applied a stretching along [001] direction. The elongation was induced by increasing the distance between the buffer layers at the wire ends in 0.1 Å steps and, subsequently, we optimized the structures within the same *ab-initio* theoretical framework. Fig. 4b shows the initial, intermediate, and final structures (before rupture, length 30.078 Å). It can easily be observed that tubular wires can not be formed as a spontaneous evolution of a 5/4 fcc rods, but this sequence reveals a very important fact: atomic planes containing 5 atoms ($3^{rd}$, $5^{th}$, $9^{th}$ and $11^{th}$ ones from the bottom in Fig. 4b, right side) seem to easily split into two planes containing 4 and 1 atoms respectively. For example, the final stacking of the first 7 layers is 5/4/4/1/4/1/4/4/5, what indeed includes 4/4 stacking sequences.

The HRTEM image interpretation also suggests that the removal of central atoms of atomic planes containing 5 atoms is the key element to generate 4/4 stacking of hollow wires. This might indicate that perhaps high stress gradients may be essential to induce the formation of hollow structures and could be one of the reasons why we did not theoretically obtain hollow tubes by a rather soft stretching of 5/4 structures. To test this hypothesis, we repeated the calculations now under the conditions of higher stress (increasing the interlayer distances between all atomic planes along the nanowires in steps of 0.5 Å), but, again, hollow tubes were not formed. Another possibility is that the shape of the basis could be important (since it affects the stress redistribution). In these terms, we must analyze the next thicker square-section Ag nanowire, which is 1.5 lattice parameter wide and is formed by the stacking of two different planes containing 8 atoms ($8_A/8_B$ stacking, the geometrical deduction of this thicker rod structure is shown in Fig. S2, Supplementary Material). Both atomic planes in the $8_A/8_B$ wire can be easily decomposed in 2 squares containing 4 atoms, and they will naturally split into a 4/4 stacking if we extend the mechanism discussed previously ($8_A/8_B$ wire, see Fig. 4c). We have then built a $8_A/8_B$ fcc wire with similar number of atoms (64 atoms in comparison with 59 for the 5/4 structures, see Fig. 4c). We then tested the low and high stress regimes using a similar procedure, starting with a relaxed $8_A/8_B$ structure and keeping frozen the end layers. The low stress elongation regime does not generate hollow tubes, but for the high stress one, we did obtain hollow structures ($4_A/4_B$ stacking) with just two steps of interlayer spacing increase (see evolution in Fig. 4c), getting a structure fully consistent with the experimentally observed ones. It is interesting to note the $4_A/4_B$ tubular region is connected to the nanowire buffer layers by two short [001] regions with a 2/2 stacking (right side of Fig. 4c); this structure has also been predicted using the geometrical Wulff construction (see Fig. S2, Supplementary Material).

We have shown that a suitable combination of time-resolved atomic resolution experiments and *ab-initio* theoretical modeling has unraveled the spontaneous formation of the smallest possible (one-lattice-parameter-wide) square metal nanotube during Ag nanowire stretching. The quantum ballistic conductance of these tubular nanowires should be ~3.6 $G_0$ ($G_0$ conductance quantum) as predicted using the methodology described in Ref [17]; this signature could be a relevant information to make this structure identifiable from electronic transport experiments. The theoretical analysis suggests that the formation of hollow structures requires a combination of minimum basis size and high gradient stress; this could explain why these structures have not been reported before, even in the theoretical simulations where low stress regimes and small structures have been the usual approach. Our results demonstrate that the proper understanding of the mechanical deformation of nanosystems requires the analysis of high symmetry metastable atomic arrangements. In fact, the revealed metastable hollow atomic structure of square cross-section fulfill a dominant energy term in nanosystems, i.e. the surface energy minimization requirement (exposing only [100] kind of facets); and, at the same time, generates a soft structure capable of absorbing a huge tensile deformation when high stress is applied.

## METHODS
### HRTEM IMAGING AND DATA TREATMENT

Silver nanowires were produced inside the HRTEM (JEM-3010 URP 300kV, 0.17 nm point resolution). Firstly, holes are opened in a self-supported metal film by focusing the electron beam and, nanometric bridges are formed between neighboring holes[18]. These constrictions evolve, elongate and break; this process is recorded with a high sensitive TV camera (Gatan 622SC). The dynamic HRTEM observations have been realized at room (~300 K) and low temperature (~150 K) using a liquid $N_2$ cooled Gatan 613-DH sample holder[19,20]. We would like to stress that the peculiar Ag nanowire structures described here are not the result of any additional special procedures to the method introduced by Takayanagi´s group[18].

It must be emphasized that Ag wires are very difficult to deal with, because the Ag melting temperature is lower than Au one; in consequence, the structural evolution is much faster. Typically, just a few frames contain all the nanowire thinning and breaking for experiments performed at room temperature[8,10]; then it is rather difficult to get a clear data on the dynamics with such limited time resolution (0.033 s). Additional complication is derived from Ag lower contrast in HRTEM due to its lower atomic number (47 versus 79 for Au). To improve this issue, we have observed the elongation process while keeping the silver nanowires at low temperature, what renders the dynamics of atom and defects propagation much lower, and better recordings can be generated. Our precedent data taken at room temperature in 2001-2002[8] did not show enough signal-noise-ratio to make the identification of the bamboo-like contrast contrast reliable.

Images have been taken at Scherzer defocus, then contrast of atomic columns can be, in a first approximation, considered proportional to the number of atoms along the observation direction. Interpretation of image contrast profiles has followed data processing as described in Ref. [10,12]. This procedure provides information on the projected structure and, by assessing the number of atoms in each column, we also get direct verification of the three dimensional atomic arrangement.

### TOTAL ENERGY CALCULATIONS AND MOLECULAR SIMULATIONS:
The applied methodology based on the DFT method was carried out using the SIESTA code[13,14] that uses norm-conserving pseudo potentials (PP) and atomic orbitals as a basis set. We have used the PP built on Troullier-Martins scheme[21,22] in a local density approximation (LDA)[23]. Firstly, we had optimized the PP to get the better lattice parameter for Ag (Ag-$a_0$) bulk. The Ag-$a_0$ was obtained with four atoms in primitive cell, for this optimization the SIESTA program allow us set free a, b, c, alpha, beta, and gamma, length and angles, respectively. The system was optimized when forces on atoms are bellow to 0.01 eV/Å. We found the Ag-$a_0$ less than 2% of difference compared with the experimental ones. Fcc and hollow nanowires were studied only in the static case not considering dynamics in a molecular dynamics framework (time dependence), a similar procedure has been used recently[23].

**References**

[1] Onishi, H., Kondo, Y. & Takayanagi, K. Quantized conductance through individual rows of suspended gold atoms. *Nature* **395**, 780-783 (1998).
[2] Yanson, A.I., Bollinger, G.R., van den Brom, H.E., Agrait, N. & van Ruitenbeek, J.M. Formation and manipulation of a metallic wire of single gold atoms. *Nature* **395**, 783-785 (1998).
[3] Rodrigues, V. & Ugarte, D. Real-time imaging of atomistic process in one-atom-thick metal junctions. *Phys. Rev. B* **63**, 073405 (2001).
[4] Kondo, Y. & Takayanagi, K. Synthesis and characterization of helical multi-shell gold nanowires. *Science* **289**, 606–608 (2000).
[5] Gülseren, O., Ercolessi, F.& Tosatti, E. Noncrystalline structures of ultrathin unsupported nanowires. *Phys. Rev. Lett.* **80**, 3775-3778 (1998).
[6] Agraït, N., Yeyati, A. L. & van Ruitenbeek, J. M. Quantum properties of atomic-sized conductors. *Phys. Rep*. **377(2–3)**, 81–279 (2003).
[7] Marks, L.D. Experimental Studies of Small-particle Structures. *Rep. Prog. Phys*. **57**, 603-649 (1994).
[8] Rodrigues, V., Bettini, J., Rocha, A. R., Rego, L. G. C. & Ugarte, D. Quantum conductance in silver nanowires: Correlation between atomic structure and transport properties. *Phys. Rev. B* **65**, 153402 (2002).
[9] Rodrigues, V., Fuhrer, T. & Ugarte, D. Signature of atomic structure in the quantum conductance of gold nanowires. *Phys. Rev. Lett.* **85**, 4124–4127 (2000).
[10] Bettini, J., Rodrigues, V., González, J. C. & Ugarte, D. Real-time atomic resolution study of metal nanowires. *Appl. Phys. A* **81**, 1513–1518 (2005).
[11] González, J.C. *et al.* Indication of unusual pentagonal structures in atomic-size Cu nanowires. *Phys. Rev. Lett.* **93**, 126103 (2004).
[12] Bettini, J. *et al.* Experimental realization of suspended atomic chains composed of different atomic species. *Nature Nanotechnol.* **1**, 182 (2006).
[13] Sanchez-Portal, D., Ordejón, P., Artacho, E. & Soler, J. M. Density-functional method for very large systems with LCAO basis sets. *Int. J. Quantum Chem.* **65**, 453–461 (1997).
[14] Soler, J.M. *et al.* The SIESTA method for ab initio order-N materials simulation. *J. Phys.: Condens. Matter* **14**, 2745-2779 (2002).
[15] Cheng, D., Kim, W.Y., Min, S.K., Nautiyal, T. & Kim, K.S. Magic structures and quantum conductance of [110] silver nanowires. *Phys. Rev. Lett.* **96**, 096104 (2006).
[16] Jia, J., Shi, D., Zhao, J. & Wang, B. Structural properties of silver nanowires from atomistic descriptions. *Phys. Rev. B* **76**, 165420 (2007).
[17] Rego, L.G.C., Rocha, A.R., Rodrigues, V. & Ugarte D. Role of structural evolution in the quantum conductance behavior of gold nanowires during stretching. *Phys. Rev. B* **67**, 165420 (2003).
[18] Kondo, Y. & Takayanagi, K. Gold nanobridge stabilized by surface structure. *Phys. Rev. Lett.* **79**, 3455-3458 (1997).

[19] Oshima, Y., Onga, A. & Takayanagi, K. Helical gold nanotube synthesized at 150 K. *Phys. Rev. Lett.* **91**, 205503 (2003).
[20] Lagos, M., Rodrigues, V. & Ugarte, D. Structural and electronic properties of atomic-size wires at low temperatures. *J. Electron Spectrosc. Rel. Phenom.* **156-158**, 20-24 (2007).
[21] Perdew, J.P. & Zunger, A. Self-interaction correction to density-functional approximaions for many-electron systems. *Phys. Rev. B* **23**, 5048-5079 (1981).
[22] Troullier, N. & Martins, J.L. Efficient Pseudopotentials for plane-wave calculations. *Phys. Rev. B* **43**, 1993-2006 (1991).
[23] Rodrigues, V., Sato, F., Galvão, D.S.& Ugarte, D. Size limit of defect formation in pyramidal Pt nanocontacts. *Phys. Rev. Lett.* **99**, 255501 (2007).

**Acknowledgements**
P.C. Silva is acknowledged for assistance during sample preparation. Supported by LNLS, FAPESP and CNPq.
Correspondence and requests for materials should be addressed to D.U.
Supplementary information accompanies this paper at www.nature.com/naturenanotechnology. Reprints and permission information is available online at http://npg.nature.com/reprintsandpermissions/.

**Author contributions**
M.J.L., J.B., V.R. and D.U. were responsible for the experimental work. F.S. and D.S.G performed the theoretical calculations. All authors discussed the results and commented on the manuscript.

**Figure Captions**

Figure 1 Elongation of a silver nanowire along [001] axis. The wire initially shows a rod-like morphology, which then gradually thins, forming an atomic chain (10.3 s) before breaking. Images were taken at ~150 K, and atomic positions appear dark in the image.

Figure 2 Structure of silver nanowires. a, Left: fcc unit cell, and the expected image contrast when a [001] nanowire of width $a_0$ is projected along [110] and [100] directions; right: analogous scheme after removing the atom located at the center of the [001] planes, generating a hollow nanowire. b, HRTEM image simulation of a hollow silver wire observed along [100] (purely geometrical non-relaxed structure), and intensity profile comparison with experiments for the two different positions indicated with an arrow in Figure 1 (profiles have been shifted vertically to aid visualization). Atomic positions appear dark in HRTEM simulation, and are represented by valleys in the intensity profile; the numbers in the profile indicate how many atoms are projected at that position.

Figure 3 Apparent nanowire axial rotation. Top: closer view of the silver wire region, revealing the image oscillating between the bamboo-like and hexagonal contrast pattern. Bottom: cross-section view of the tubular nanowire accounting for HRTEM images (microscope electron beam is considered in the vertical direction and, the corresponding crystallographic orientations are indicated in the lower part of the figure). The observed structural fluctuation is attributed to a radial movement of silver atoms of the tubular nanowire (one plane contracts while the other expands) and it is visible in HRTEM images as an apparent nanowire rotation.

Figure 4 Theoretical analysis of the derived atomic wire structures. a, Schema of the relaxed atomic arrangement of a one-lattice-parameter wide wire formed by 13 stacked atomic layers; left and right side are fcc and hollow wire, respectively. b, Low stress elongation of fcc 5/4 structure, which does not lead to the formation of tubular wires. c, Slightly thicker (1.5 $a_0$) fcc wire is generated by the stacking of two planes containing 8 atoms each ($8_A/8_B$ stacking), and a high stress elongation generates the $4_A/4_B$ stacking of hollow silver nanowires (see text for further explanation). To aid interpretation, all atomic planes containing 4 atoms have been coloured black.

**Table 1.**

Formation energy per atom of fcc and tubular one-lattice-parameter-thick [001] silver nanowires of different length. The 4$^{th}$ column indicates how many atoms were extracted from the fcc wire axis to generate the corresponding tubular one.

| Number of stacked atomic layers | Fcc wires | | Tubular wires | |
|---|---|---|---|---|
| | Number of Atoms | Energy (eV) | Number of vacancies | ($E_{tub}$ - $E_{fcc}$) (eV) |
| 5 | 23 | -4.035299 | 1 | 0.078759 |
| 7 | 32 | -4.195817 | 2 | 0.098302 |
| 9 | 41 | -4.278174 | 3 | 0.139482 |
| 11 | 50 | -4.337848 | 4 | 0.163262 |
| 13 | 59 | -4.344577 | 5 | 0.143624 |

Figure 1

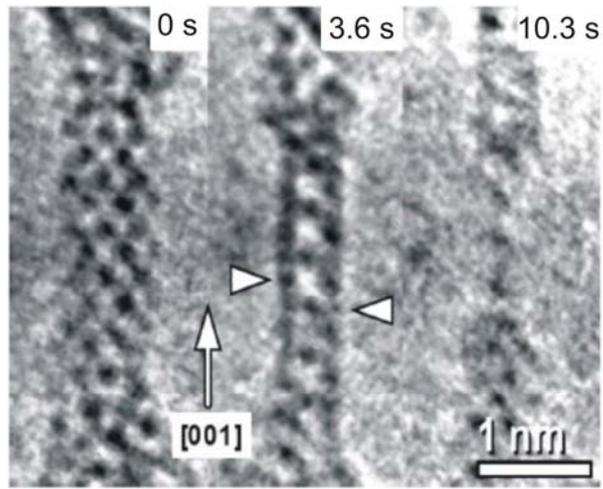

Figure 2

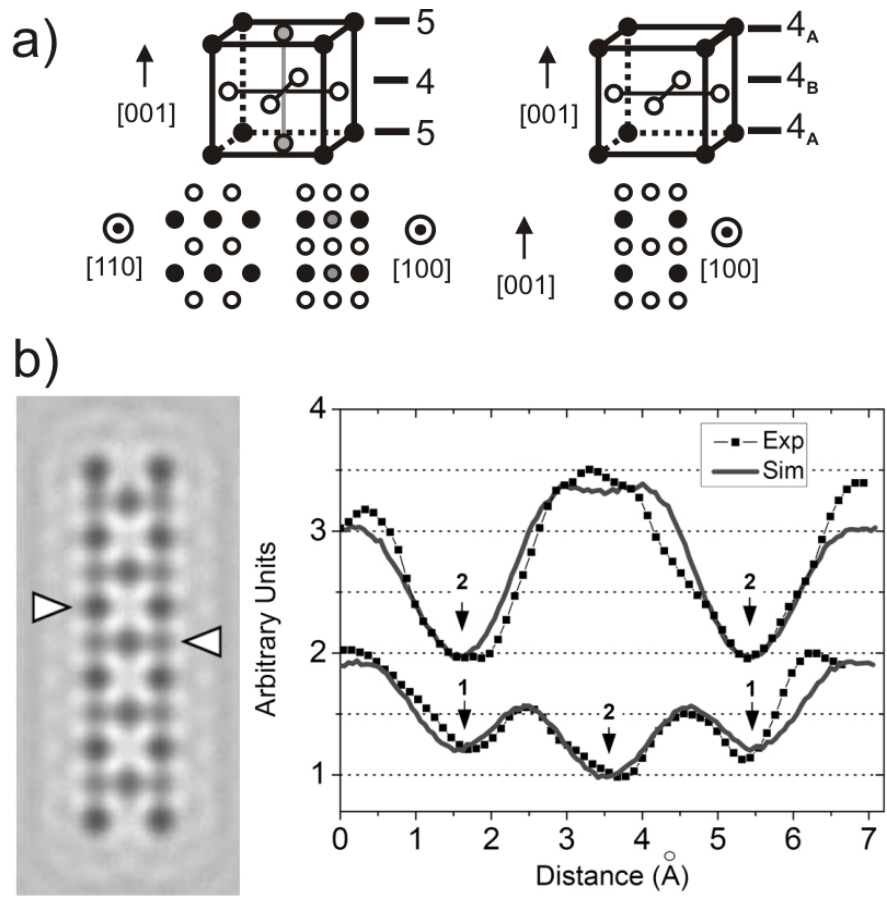

Figure 3

Figure 4

# Supplementary Information

## Observation of the Smallest Metal Nanotube With Square Cross-Section

M.J. Lagos[1,2], F. Sato[2], J. Bettini[1], V. Rodrigues[2], D.S. Galvão[2], D. Ugarte [1,2]*
[1]Laboratório Nacional de Luz Síncrotron, C.P. 6192,
13083-970 Campinas SP, Brazil.
[2]Instituto de Física "Gleb Wataghin", Universidade Estadual de Campinas-UNICAMP,
13083-970 Campinas SP, Brazil.
*To whom correspondence should be addressed. E-mail: dmugarte@ifi.unicamp.br

### Includes

| | | |
|---|---|---|
| Videos: | 01 | (avi format, 6 Mb) |
| | 02 | (avi format, 1Mb) |
| Figures | S1 | |
| | S2 | |

Fig. S1

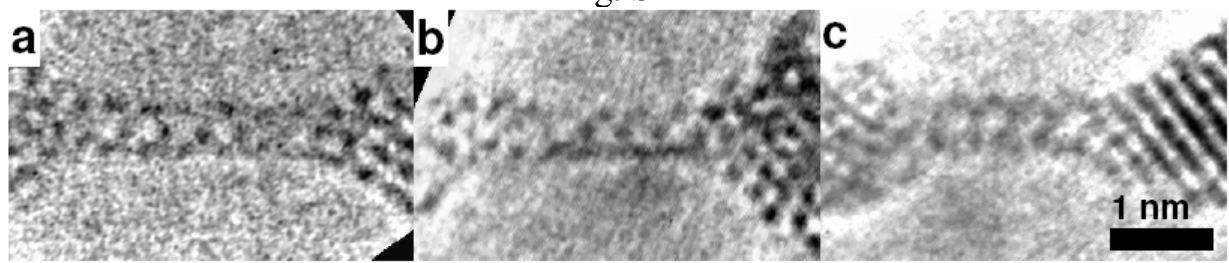

Fig. S2

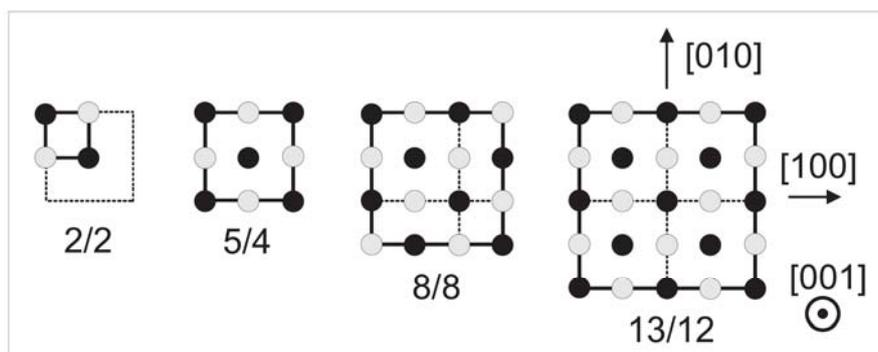

**Video Captions:**

Video 01: Elongation and rupture of a rod-like Ag NW under tensile stress along the [001] crystalline direction.

Video 02: Schematic representation of a hollow NW structural fluctuation that should appear as NW apparent rotation along it axis when observed with a HRTEM experiment.

**Figures Captions:**

Fig S1. Snapshots extracted from video recording of [001] Ag rods displaying bamboo-like contrast. Images (a-b) and (c) were obtained at ~300 and 150 K, respectively. Atomic positions appear dark.

Fig. S2. Cross-section view of [001] Ag rod-like wires (0.5 to 2 $a_0$ in width) derived by the application of the Wulff construction. Surface energy minimization predicts [100] and [010] facets at the wire surface, then the wire cross-section should be always square. The numbers below the structures indicate the corresponding stacking sequence. Black and grey colors are used to indicate the atom location in different stacked planes. Note that the 8/8 structure (1.5 $a_0$ wide) represents the low surface energy [001] Ag wire, which is the next thicker one beyond the 5/4 wire.